\documentclass{svproc}
\usepackage{url}
\usepackage{wrapfig}
\usepackage{lipsum} 
\usepackage{graphicx}
\usepackage[T1]{fontenc}
\usepackage[utf8]{inputenc}
\usepackage{lmodern}
\usepackage{wrapfig}
\usepackage{chngcntr}
\usepackage{pdflscape}
\usepackage[lastpage,user]{zref}
\usepackage{lipsum}
\usepackage{graphicx}

\mainmatter              
\title{Rare decays: A window on new
physics }
\titlerunning{Rare decays: A window on new
physics}  
\author{N. Dash\inst{}} 

\authorrunning{N. Dash}

\institute{Indian Institute of Technology Bhubaneswar, INDIA \\\email{E-mail: nd11@iitbbs.ac.in}}

\begin{document}

\maketitle              

\begin{abstract}
We report on the extrapolated measurements related to the flavour-changing-neutral-current(FCNC) $b\rightarrow s$ and $b\rightarrow d$ transitions with the Belle II data. The branching fraction(BF) and raw asymmetry measurements of the exclusive decay $\bar{B}\rightarrow X_{q}\gamma$, time-dependent $CP$ asymmetry in the transition $b\rightarrow s\gamma$, angular analysis of $B\rightarrow K^*ll$ and $B\rightarrow X_sll$ are discussed. We also report on the searches for the decay $B\rightarrow h\nu\nu$. Most of these analyses are extrapolated with 5 and 50 ab$^{-1}$ Belle II data. 
\end{abstract}

\section{Introduction}
B-factories had 10 years of the successful operational period and accumulated 1.5 ab$^{-1}$ data (1.25 $\times 10^9$ $B\bar{B}$pairs). The major achievements of Belle include observation of $CP$ violation in $B$ meson system and confirmation of CKM picture, first
evidence for mixing in the $D$ meson system, first evidence for exotic states X(3872) and so on. Belle II, as a next generation flavor factory, aims to search for New Physics (NP) in the flavour sector and to further reveal the nature of QCD. Belle II is expected to gather 50 ab$^{-1}$ of data in $e^+e^-$ collisions by 2025~\cite{b2tdr}.
\\\\FCNC $b\rightarrow s$ and $b\rightarrow d$ processes continue to be of great importance to precision flavor physics (FP).
Final states having color singlet leptons and photons are both theoretically and experimentally
clean. Also, radiative and electroweak (EW) penguin $B$ decays are an ideal place to search for NP.
Belle II physics program in this area will focus on the process such as inclusive measurements of
$B\rightarrow X_{s,d}\gamma$, $B\rightarrow X_{s,d}ll$ as well as decay $B\rightarrow K^{(*)}\nu\nu$~\cite{b2tip}. 
The fully-inclusive measurements with final states containing pairs of photons, neutrinos or taus are possible at Belle II. It will provide an independent test of the anomalies recently uncovered by the LHCb and Belle experiments in the angular analysis of $B\rightarrow K^*\mu^+\mu^-$ and in the determination of $R(K)$.
\section{$\bar{B}\rightarrow X_{q}\gamma$ decay}
\subsection{Branching Fraction}
\begin{figure}
 \centering
 \includegraphics[width=4cm,height=4.3cm]{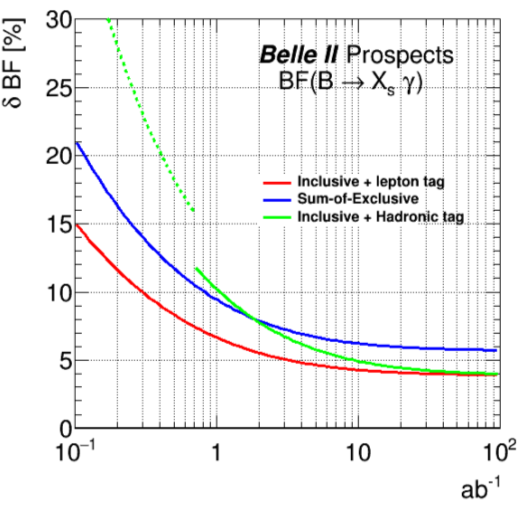} 
 \includegraphics[width=4cm,height=4.2cm]{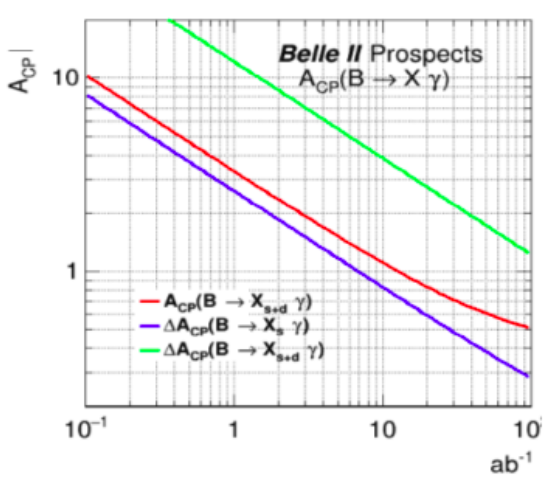} 
 \includegraphics[width=4cm,height=4.1cm]{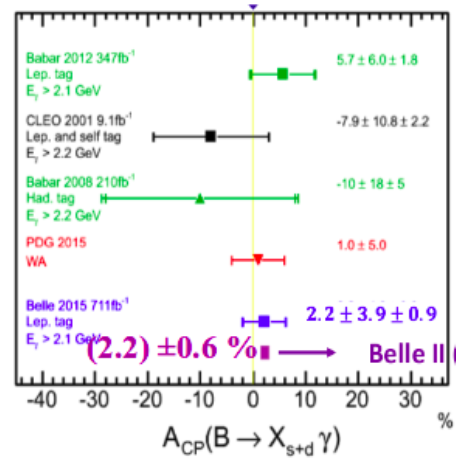} 
 \\
 \caption{Belle II exceptions on BF and $CP$ uncertainty of $B\rightarrow X_{s}\gamma$ decays.}
 \label{fig:bf}
 \end{figure}

The inclusive $\bar{B}\rightarrow X_{s,d}\gamma$ decays provide important constraints on masses and interactions of
many possible beyond Standard Mechanism (BSM) scenarios such as models with extended Higgs sector or super-symmetric (SUSY). Also, it is sensitive to |$C_7$| co-efficient.  Precise SM prediction~\cite{th} is available (for the $CP$ and isospin asymmetry (IA) branching ratios) for gamma threshold energy ($E_{\gamma}$) greater than 1.6 GeV are as :
\begin{equation}
Br_{s\gamma} = (3.36\pm0.23)10^{-4}\hspace{30pt}
Br_{d\gamma} = (1.73^{+0.12}_{-0.22})10^{-5},
\end{equation} 
and experimental obtained results~\cite{ex} are as :
 \begin{equation}
 Br_{s\gamma} = (3.27\pm 0.14)10^{-4}\hspace{30pt}
 Br_{d\gamma} = (1.41\pm 0.57)10^{-5}.
 \end{equation} 
 Though experiment and theory are consistent and put a strong limit on NP, experimental measurements are systematically dominated. The  Belle  measurement on fully inclusive method is systematic dominated~\cite{409}. This systematics can be reduced with large Belle II data samples. The extrapolated Belle II prospects on BF will be 3.9\% with full 50 ab$^{-1}$ as shown in  Fig.~\ref{fig:bf}~\cite{b2tip}. 
 
\subsection{Rate Asymmetry}
In addition to BFs, asymmetry in decay rates (IA and $CP$ asymmetry) are also sensitive to BSM contributions. The SM predictions on $A_{CP}$ of the decays $B\rightarrow X_{s,d}\gamma$ are given in the Ref.~\cite{th1,th2}.
The direct $CP$ asymmetry and their IA are defined as:
\begin{equation}
A_{CP}=\frac{\Gamma(\bar{B}\rightarrow X_s\gamma)-\Gamma(B\rightarrow X_{\bar{s}}\gamma)}{\Gamma(\bar{B}\rightarrow X_s\gamma)+\Gamma(B\rightarrow X_{\bar{s}}\gamma)},
\end{equation}
\begin{equation}
\Delta A_{CP}(B\rightarrow
X_q\gamma)=A_{CP}(B^+\rightarrow X_q^+\gamma)-A_{CP}(B^0\rightarrow X_q^0\gamma).
\end{equation}
The existing measurements of $A_{CP}$ are measured by BaBar~\cite{461} and Belle~\cite{462} by using the sum-of-exclusive method and BaBar also measured $\Delta A_{CP}$ for fully inclusive method~\cite{405}. Belle II can be measured both $A_{CP}$ and $\Delta A_{CP}$ using the same technique as BaBar, yet with a much larger data set. A reduction of the systematic uncertainties is therefore crucial at Belle II. The obtained previous results and expected Belle II results are shown in Fig.~\ref{fig:bf}~\cite{b2tip}.
\section{$b\rightarrow s\gamma$ transitions (Time dependent $CP$ asymmetry)}
The observable IA, sensitive to BSM, can be defined as :
\begin{equation}
a_I^{0-} = \frac{c_V^2\Gamma(\bar{B^0}\rightarrow \bar{V^0}\gamma-\Gamma(B^-\rightarrow V^-\gamma))}{c_V^2\Gamma(\bar{B^0}\rightarrow \bar{V^0}\gamma+\Gamma(B^-\rightarrow V^-\gamma))}
\end{equation}  
where $c_{\rho^0}=\sqrt{2}$ and $c_{K^{*0}}=1$ are isospin-symmetry factors~\cite{b2tip}. To accumulate more statistics one can define CP-averaged IAs through $\bar{a}_I$ = $\bar{a}_I^{0-}$+ $a_I^{0+}$/2. The most up-to-date theoretical predictions~\cite{18} for the IAs are 
\begin{equation}
\bar{a}_{I}^{SM}(K^*\gamma) = (4.9 \pm 2.6)\%,\hspace{10pt}
\bar{a}_{I}^{SM}(\rho\gamma) = (5.2 \pm 2.8)\%
\end{equation}
and these are consistent with the HFLAV average~\cite{hflav} which are as: 
\begin{equation}
\bar{a}_{I}^{exp}(K^*\gamma) = (5.2\pm 2.6)\%,\hspace{10pt}
\bar{a}_{I}^{exp}(\rho\gamma) = (30^{-13}_{+16})\%.
\end{equation}
The observable,  
\begin{equation}
1-\delta_{aI} = \frac{\bar{a}_{I}(\rho\gamma)}{\bar{a}_{I}(K^*\gamma)}\sqrt{\frac{\bar{\Gamma}(B\rightarrow \rho\gamma)}{\bar{\Gamma}(B\rightarrow K^*\gamma)}}\bigg|\frac{V_{ts}}{V_{td}}\bigg|
\end{equation}
where $\delta_{aI}$ is close to zero, and the quantity (1 $-$ $\delta_{aI}^{SM}$)= 0.90 $\pm$ 0.11 shows a reduced uncertainty with respect to the individual CP-averaged IAs. The experimental average $\delta_{aI}^{exp}$ = $-$4.0 $\pm$ 3.5 can be improved at Belle II through more statistics as well as taking into account experimental correlations. In the SM, expected mixing-induced $CP$ asymmetries ($S$) are as follows:
\begin{equation}
S_{K^*(K^0_S\pi^0)\gamma} = -2\frac{m_s}{m_b}\sin2\phi_1 = a few \%,\hspace{10pt} S_{\rho(\pi^+\pi^-)\gamma} = 0. 
\end{equation}
The expected uncertainties on $S$ are 0.03\% (0.09\%) and 0.06\% (0.19\%) for the decay modes $K^{*}(K_S^0\pi^0)\gamma$ and $\rho^0(\pi^+\pi^-)\gamma$ respectively with 50 (5) ab$^{-1}$ Belle II data sample~\cite{b2tip}. 
\section{Measurements of $R_K$, $R_{K^*}$, and $R_{X_s}$}
The decay $B \rightarrow K^{*}ll$ proceeds via one loop diagram and lepton universality holds in SM. Interestingly, in the recent years several measurements have shown possible deviations from the SM
for this decay~\cite{sau}. The Lepton Flavor Universality (LFU) ratios $R_K(R_{K^*})$ are defined as the ratios $\frac{\Gamma[B \rightarrow K(K^{*})\mu\mu]}{\Gamma[B \rightarrow K(K^{*})ee]}$ and $R_{X_s}$ as $\frac{\Gamma[B \rightarrow X_s\mu\mu]}{\Gamma[B \rightarrow X_see]}$. The electron mode is challenging at LHCb, especially for high $q^2$ (invariant mass squared of the two leptons) region but Belle II is having similar efficiency for electron and muon modes. Thus, the measurements at both low and high $q^2$ regions and the ratios $R_K$, $R_{K^*}$, and $R_{X_s}$ are possible at Belle II.
 \begin{figure}
 \centering
 \includegraphics[width=12.5cm,height=4cm]{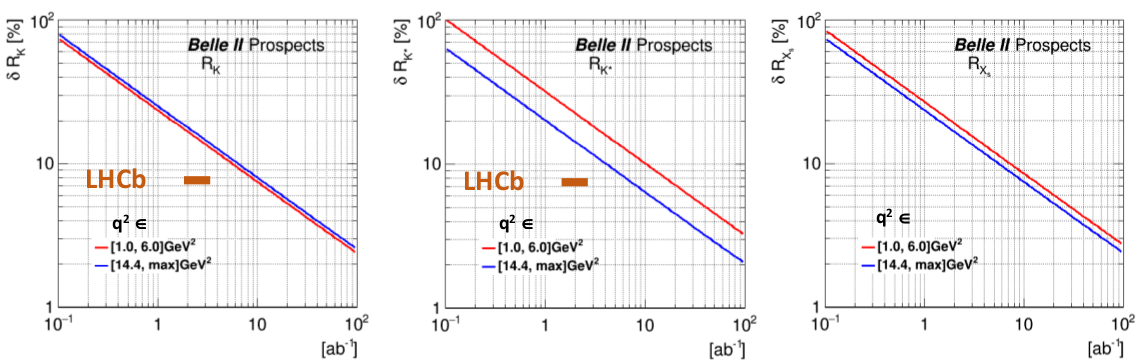} \\
 \caption{Expected uncertainty of $R$ measurement at Belle II.}
 \label{fig:rk}
 \end{figure}
\begin{figure}[h!]
 \centering
 \includegraphics[width=6cm,height=4cm]{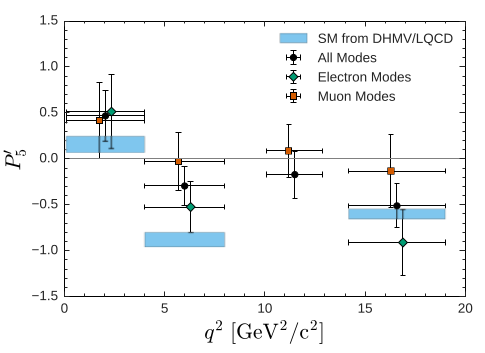} 
\includegraphics[width=6cm,height=4cm]{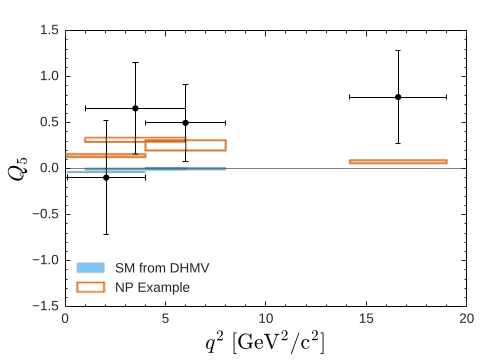}  
 \caption{(left) $P_5$observables for combined, electron and muon modes and (right) $Q_5$ observables compared with SM and NP scenario.} 
 \label{fig:p5}
 \end{figure}
   \begin{figure}[h!]
   \centering
   \includegraphics[width=12cm,height=4cm]{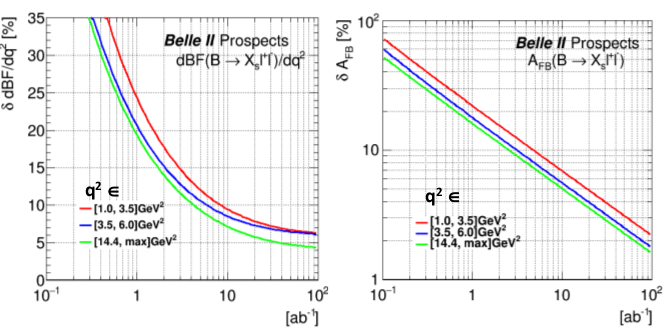} \\
   \caption{Belle II exceptions on BF and $A_{FB}$ uncertainty of $B\rightarrow X_{s}ll$ decays.}
   \label{fig:afb}
   \end{figure}
An additional $q^2$ $\epsilon$ (1.0, 6.0) GeV$^2$/c$^2$ bin is considered, which is favored for theoretical predictions~\cite{43}. To maximize the potency of limited statistics, a data-transformation technique is utilized~\cite{48,49}. The result is shown in Fig.~\ref{fig:p5}, where it is also compared with SM predictions~\cite{50,51}. The largest deviation is 2.6$\sigma$ , observed in $q^2$ $\epsilon$ (4.0, 8.0) GeV$^2$ /c$^2$ bin of $P_5$ for the muon mode~\cite{simon}. This tension is coincidental to the $P_5$ anomaly earlier reported by LHCb~\cite{47,48}. In the same region the electron modes deviate by 1.3$\sigma$ and the combination deviates by 2.5$\sigma$.
The observables $P_5$ and $Q_5$ are presented in Fig.~\ref{fig:p5}, where they are compared with SM and NP scenario~\cite{46}. The results show no significant deviation from zero. A global fit performed including these measurements~\cite{simon} suggests for lepton-universality violation~\cite{52}. Belle II and LHCb will be comparable for this $b\rightarrow sll$ process. The Belle II projection for $P_5$ anomaly for different $q^2$ regions are listed in Table\ref{tab:p5}. In the low $q^2$ region the uncertainty will be 4\% with 50 ab$^{-1}$~\cite{b2tip} which will be comparable with LHCb 22 fb$^{-1}$ result. The Belle II excepted uncertainty on BF and $A_{FB}$ uncertainty of $B\rightarrow X_{s}ll$ decays are shown in  Fig.~\ref{fig:afb}~\cite{b2tip}.


\begin{table}[h!]
 \centering
 	\scalebox{0.9}{
 	\begin{tabular}{c c c } \hline
 	$q^2$ (GeV$^2$/c$^4$)  &  Belle & Belle II  \\ \hline
 0.1-4      &  0.42 & 0.06 \\
 4-8  & 0.28 & 0.04\\
 10.09-12 & 0.34 & 0.05\\
 14.18-19 & 0.25 & 0.03\\\hline
 	\end{tabular} }
 	\caption{Expected uncertainty on $P_5$ at Belle II in bins of $q^2$.}
 	\label{tab:p5}
 	\end{table} 
\section{Search for $b\rightarrow s\nu\nu$ transitions}
The decays $B \rightarrow h\nu\nu$ (where $h$ refers to $K^+$, $K_S ^0$ , $K^{*+}$ , $K^{*0}$ , $\pi^+$ , $\pi^0$ , $\rho^+$ or $\rho^0$ ~\cite{K*}) are theoretically clean due to the exchange of a $Z$ boson alone, in comparison to other $b \rightarrow s$ transitions where the
virtual photon also contributes~\cite{hnunuth}. The SM predicted BF is given in the Ref.~\cite{hnunuth}. Previously, the decays $B \rightarrow h\nu\nu$ have been searched in Belle utilizing the hadronic tag method~\cite{55} and in BaBar using both hadronic~\cite{56} and semi-leptonic (SL)  tag~\cite{57}. The recent Belle analysis updated this measurement with SL tag method and set most stringent limits till date in most channels~\cite{hnunu}. 
\begin{figure}[h!]
 \centering
 \includegraphics[width=6.5cm,height=4.75cm]{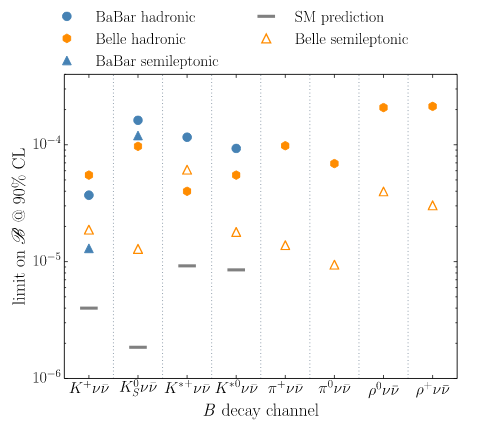} \\
 \caption{Belle II exceptions on BF uncertainty of $B\rightarrow h\nu\nu$ decays.}
 \label{fig:hnunu}
 \end{figure}The SM predicted the $K^{(*)}$ mode is a golden channel for Belle II and extrapolated by assuming Belle hadronic and SL tag analyses as 100\%. The precision on the BF of few decay modes are listed in Table~\ref{tab:hnunu1} at 50 ab$^{-1}$ Belle II data set. The BF of $B\rightarrow K(*)\nu\nu$ is measurable at Belle II with about 10\% uncertainty~\cite{b2tip}.
 \begin{table}[h!]
 \centering
 	\scalebox{1.0}{
 	\begin{tabular}{c c c } \hline
 	Mode  &  stat. only & Total  \\ \hline
 $B^+ \rightarrow K^+\nu\nu$      &  10\% & 11\% \\
 $B^+ \rightarrow K^{*+}\nu\nu$ & 8\% & 9\% \\
 $B^+ \rightarrow K^{*0}\nu\nu$ & 8\% & 10\%\\\hline
 	\end{tabular} }
 	\caption{Expected uncertainty on $K$ modes at Belle II.}
 	\label{tab:hnunu1}
 	\end{table}



\begin{thebibliography}{99}
\bibitem{b2tdr} T. Abe \textit{et al.} (Belle II Collaboration), Belle II Technical Design Report, arXiv:1011.0352.
\bibitem{b2tip} E. Kou, P. Urquijo, The Belle II Collaboration, and The B2TiP theory community, arXiv:1808.10567.
\bibitem{th} M. Misiak \textit{et al.} Phys. Rev. Lett. \textbf{114}, 221801 (2015).
\bibitem{ex} Michal Czakon \textit{et al.} JHEP, 04, \textbf{168} (2015).
\bibitem{409} T. Saito \textit{et al.}, Belle Collaboration, Phys. Rev.,
D \textbf{91} (5), 052004 (2015), arXiv:1411.7198.
\bibitem{th1} T. Hurth, E. Lunghiand W. Porod, Nucl.Phys. B \textbf{704} 56–74 (2005).
\bibitem{th2} M. Benzke \textit{et al.} Phys. Rev. Lett. \textbf{106} 141801 (2011).
\bibitem{461} B. Aubert \textit{et al.}, BaBar Collaboration, Phys. Rev., D \textbf{72}, 052004 (2005), arXiv:hep-ex/0508004.
\bibitem{462} S. Watanuki \textit{et al.}, Belle Collaboration, arXiv:1807.04236 (2018), BELLE-CONF-1801.
\bibitem{405} B. Aubert \textit{et al.}, BaBar Collaboration, Phys. Rev., D \textbf{77}, 051103 (2008), arXiv:0711.4889.
\bibitem{18} J. Lyon and R. Zwicky, Phys. Rev. D \textbf{88}, 094004 (2013).
\bibitem{hflav}bhttp://www.slac.stanford.edu/xorg/hflav/triangle/summer2017/index.shtml.
\bibitem{sau} W. Altmannshofer and D. M. Straub, Eur. Phys. J. C\textbf{75}, 382 (2015).
\bibitem{43} W. Altmannshofer, P. Ball, A. Bharucha, A. J. Buras, D. M. Straub, and M. Wick, JHEP \textbf{01}, 019
(2009).
\bibitem{48} R. Aaij \textit{et al.} (LHCb Collaboration), Phys. Rev. Lett. \textbf{111}, 191801 (2013).
\bibitem{49} M. D. Cian, Ph.D. thesis, University of Zurich (2013).
\bibitem{50} S. Descotes-Genon, \textit{et.el.}, JHEP \textbf{12}, 125 (2014). 
\bibitem{51} R. R. Horgan, \textit{et.el.},  PoS LATTICE2014,
 \textbf{372} (2015). 
 \bibitem{simon} S. Wehle et al. (Belle Collaboration), Phys. Rev. Lett. \textbf{118}, 111801 (2017).
\bibitem{47} R. Aaij et al. (LHCb Collaboration), JHEP \textbf{02}, 104 (2016).
\bibitem{46} B. Capdevila, \textit{et.el.}, JHEP \textbf{10}, 075 (2016). 
\bibitem{52} B. Capdevila, \textit{et.el.}, arXiv:1704.05340 [hep-ph]. 
\bibitem{K*} The $K^{*0} (892)$ is denoted as $K^{*0}$.
\bibitem{hnunuth} Andrzej J. Buras \textit{et.el.} JHEP 02 \textbf{184}, 2015, arXiv:1409.4557 [hep-ph]. 
\bibitem{55} O. Lutz \textit{et.el.} (Belle Collaboration), Phys. Rev. D \textbf{87}, 111103 (2013).
\bibitem{56} J.P. Lees \textit{et.el.} (BaBar Collaboration), Phys. Rev. D \textbf{87}, 112005 (2013).
\bibitem{57} J.P. Lees \textit{et.el.} (BaBar Collaboration), Phys. Rev. D \textbf{82}, 112002 (2010). 
\bibitem{hnunu} J. Grygier \textit{et.el.} (Belle Collaboration), Phys. Rev. D \textbf{96}, 091101(R) (2017).

\end{thebibliography}
\end{document}